\documentclass[aps,pra,reprint,groupedaddress,superscriptaddress,a4paper]{revtex4-1}
\usepackage[utf8]{inputenc}
\usepackage[normalem]{ulem}
\usepackage{graphicx}  
\usepackage{enumerate}
\usepackage{amssymb}   
\usepackage{amsmath}   
\usepackage{amsthm}   
\usepackage{comment} 
\usepackage{color}
\usepackage[colorlinks = true,
            linkcolor = blue,
            urlcolor  = blue,
            citecolor = blue,
            anchorcolor = blue]{hyperref}

\usepackage{verbatim}

\usepackage{newtxtext} 
\usepackage{newtxmath} 

\def \Log {\operatorname{Log}}

\def \Arg  {\operatorname{Arg}}

\def \plus {+}
\def \minus {-}

\def \Re {\operatorname{Re}}
\def \Im {\operatorname{Im}}

\begin{document}

\title{How vortex bound states affect the Hall conductivity of a chiral \texorpdfstring{$\boldsymbol{p\pm i p}$}{p±ip} superconductor}

\author{Daniel Ariad}
\email[]{daniel@ariad.org}
\affiliation{Department of Physics, Ben-Gurion University of the Negev, Beer-Sheva 8410501, Israel}

\author{Yshai Avishai}
\affiliation{Department of Physics, Ben-Gurion University of the Negev, Beer-Sheva 8410501, Israel}
\affiliation{NYU-Shanghai University, 1555 Century Avenue, Pudong, Shanghai 200122, China }
\affiliation{Yukawa Institute for Theoretical Physics, Kyoto University, Kyoto 606-8502, Japan}

\author{Eytan Grosfeld}
\affiliation{Department of Physics, Ben-Gurion University of the Negev, Beer-Sheva 8410501, Israel}

\begin{abstract}
The physics of a planar chiral $p\pm i p$ superconductor is studied for various vortex configurations. The occurrence of vortex quasiparticle bound states is exposed together with their ensuing collective properties, such as subgap bands induced by intervortex tunneling. A general method to diagonalize the Hamiltonian of a superconductor in the presence of a vortex lattice is developed that employs only smooth gauge transformations. It renders the Hamiltonian to be periodic (thus allowing the use of the Bloch theorem) and enables the treatment of systems with vortices of finite radii. The pertinent anomalous charge response $c_{xy}$ is calculated (using the Streda formula) and reveals that it contains a quantized contribution. This is attributed to the response to the nucleation of vortices from which we deduce the system's quantum phase.
\end{abstract}
\maketitle

\section{Introduction}
Measurement of the polar Kerr effect (PKE) in the superconducting state of Sr$_2$RuO$_4$ indicates the presence of time-reversal symmetry breaking \cite{jing2006Hhigh,kapitulnik2009polar}. However, so far no quantitative agreement has been established between theoretical and experimental values of the Kerr angle \cite{goryo2008impurity,lutchyn2008gauge,lutchyn2009frequency,taylor2012intrinsic,gradhand2013kerr,gradhand2015polar}. The latter is proportional to the Hall conductivity, which in turn is proportional to the anomalous charge response $c_{xy}$ \cite{roy2008collective}. The quantity $c_{xy}$ is finite only in a chiral superconductor \cite{read2000paired,horovitz2003superconductors}, so the measurement of the PKE provided some of the first evidence for the $p\pm ip$ nature of the order parameter of Sr$_2$RuO$_4$.

\begin{figure}[!b] 
	\centering
	\includegraphics[trim={0 0 0 0},clip,width=0.475\textwidth]{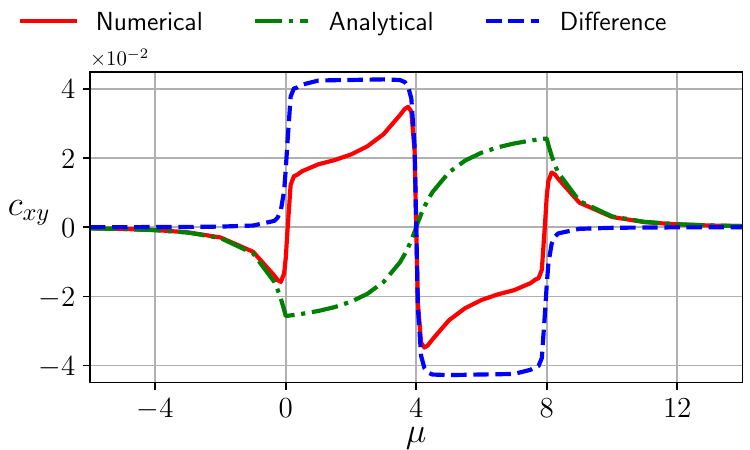}		
	\caption{Average anomalous charge response $c_{xy}$ vs chemical potential $\mu$ for a planar $p$-wave superconductor. The result of a modified Streda formula (Numerical) is compared with the prediction of the effective low-energy theory of a $p$-wave superconductor (Analytical). Here $t = |\Delta| = 1$ and $\xi = 2.5$. In addition, the magnetic unit cell contains $40 \times 41$ sites and two vortices that are pinned on its diagonal, partitioning it in a ratio of 1:2:1. \label{fig:main}}
\end{figure}

In this paper, we calculate $c_{xy}$ at zero magnetic field and zero vorticity using a modified Streda formula and show that $c_{xy}$ is a sum of two contributions, one which is nonuniversal, and the other equals $\kappa/8\pi$, where $\kappa$ is the Chern number of the superconductor, as depicted in Fig.~\ref{fig:main}. An important insight gained thereby is that an accurate evaluation of $c_{xy}$ requires the knowledge of the charge response to the application of a weak magnetic field and a compensating vortex pair as dictated by imposing periodic boundary conditions (PBCs). This is equivalent to elucidation of the charge response following a chirality flip of the superconductor. Eventually, however, the effect of vortices characteristics (such as their positions as well as their detailed structures) on  $c_{xy}$ is minor, and our main results appear to be universal. Once $c_{xy}$ is elucidated, the Hall conductivity at a zero magnetic field and vorticity can be extracted from $c_{xy}$ using a standard procedure \cite{horovitz2003superconductors,roy2008collective}, and that has bearing on the experimentally measured PKE.

In order to substantiate our main result, we need to consider the response of the superconductor to the insertion of a single Dirac flux quanta ($\Phi=h/e$) and compensating pair of vortices. Due to the PBCs imposed on the system when employing the Streda formula, it is natural to solve an equivalent problem for a system composed of many copies of the (originally finite) system, which maps onto an infinite superconductor in the presence of a periodic vortex lattice. The vortices are assumed to have finite radii, thus enabling us to explore the possible dependence of $c_{xy}$ on the presence of vortex bound states.  

A natural framework for studying the physics of a periodic vortex lattice is to employ Bloch's theorem. However, this procedure is hindered by the fact that the vector potential and the phase of the order parameter are not independently periodic over the magnetic unit cell (MUC). One may try to apply a gauge transformation to combine the two into a single field which is proportional to the supercurrent. As the latter is periodic in the lattice, Bloch theorem can be employed. However, since the gauge transformation is singular in the presence of vortices, this procedure introduces spurious magnetic fields in the center of the vortices. These spurious fields either break particle-hole symmetry or introduce branch-cuts, originating from the vortex centers, that lead to numerous technical obstacles \cite{vishwanath2002dirac,ariad2015effective,liu2015electronic,murray2015majorana,cvetkovic2015berry}.

To circumvent these obstacles, we develop an algorithm to perform an efficient exact diagonalization of the Bogoliubov-de Gennes (BdG) Hamiltonian for an infinite two-dimensional (2D) vortex lattice in a general tight-binding model, that completely avoids the use of singular gauge transformations. Instead, a smooth gauge transformation is employed, that renders both the order parameter and the hopping amplitudes to be independently periodic on the lattice sites. 

\section{Generalities}
It is our perception that the algorithm developed here for the diagonalization of the Hamiltonian  is not just a numerical trick,  but rather, it meticulously exploits the pertinent physical concepts. Thus, it is worthwhile to illuminate its construction step by step right at the onset. First, we derive an exact expression for the phase of the order parameter by summation over vortices in an ordered array in a superconductor. Second, we transform to another gauge that allows for simultaneously taking the superconducting phase function and the Peierls phases to be periodic functions (mod $2\pi$) without sacrificing any of their properties. Third, we introduce a new gauge for the vector potential, which we dub the ``almost anti-symmetric gauge (AAG),'' which allows accessing, in a system with PBCs, the highest resolution for its magnetic-field dependence. Fourth, we diagonalize the Hamiltonian in a single unit cell under varying boundary conditions per the Bloch theorem, i.e., for different values of the lattice momentum. Thus we extract both the full spectrum of the Hamiltonian and its wave functions.

\section{Hamiltonian and order parameter}  For spin-$1/2$ fermions (spin projection $s=\uparrow, \downarrow$), the BdG Hamiltonian in its tight binding form (taking $\hbar=c=e=1$) consists of three terms $\hat{H}=\hat{T}+\hat{\Delta}-(\mu-4t) \hat{N}$.  The hopping term reads 
\begin{eqnarray}
	\hat{T}=-t \sum_{\boldsymbol{r},s,i} e^{i\int_{\boldsymbol{r}}^{\boldsymbol{r}+\boldsymbol{a}_i} \mathbf{A}\cdot d\boldsymbol{\ell}}\psi^\dag_{\boldsymbol{r}+\boldsymbol{a}_i,s}\psi_{\boldsymbol{r},s}+\mbox{H.c.}
\end{eqnarray}
The pairing term for an $s$-wave superconductor is as follows:
\begin{equation}
\hat{\Delta}_{s\text{-wave}}= \sum_{\boldsymbol{r}} \Delta(\mathbf{r})  \psi^\dag_{\boldsymbol{r}\uparrow}\psi^\dagger_{\boldsymbol{r}\downarrow}+\mbox{H.c.}, 
\end{equation}
where $\Delta(\boldsymbol{r}) = \Delta_0(\boldsymbol{r}) e^{i\Theta(\boldsymbol{r})}$ with $\Delta_0(\boldsymbol{r})$, $\Theta(\boldsymbol{r})$ as real scalar fields and $\boldsymbol{a}_i=a_i\hat{\boldsymbol{\tau}}_i$ (with $i=1,2$) are the lattice vectors. For spinless fermions, we omit one spin component from the hopping term and take the lowest angular momentum $p$-wave pairing,
\begin{equation}\begin{aligned}
	\hat{\Delta}_{p\text{-wave}}= \sum_{\boldsymbol{r},i} \Delta_{p\pm ip}(\boldsymbol{r},\boldsymbol{a}_i) \psi^\dag_{\boldsymbol{r}\downarrow}\psi^\dagger_{\boldsymbol{r}+\boldsymbol{a}_i\downarrow}+\mbox{H.c.},
\end{aligned}\end{equation}
where $\Delta_{p\pm ip}(\boldsymbol{r},\boldsymbol{a})=\Delta_0(\boldsymbol{r}) e^{\pm i \mathrm{Arg}(\boldsymbol{a})}e^{i\Theta(\boldsymbol{r})}e^{\frac{i}{2}\int_\mathbf{r}^{\mathbf{r}+\boldsymbol{a}}\nabla\Theta\cdot \mathrm{d}\boldsymbol{\ell}}$ and $\Arg(\boldsymbol{r})=\Arg(x+iy)$. The superconducting order parameter is defined in such a way that the $\mathbb{U}(1)$ gauge invariance is respected \cite{ariad2015effective}.

We recall that vortices are encoded as nodes of the order parameter, characterized by a finite quantized winding number of the phase $\Theta(\boldsymbol{r})$ \cite{mermin1979topological}. In order to form a vortex lattice we tile the plane with a MUC. The MUC is chosen to enclose an even number of vortices. Thus, each vortex within the MUC constitutes a sublattice. The superconducting phase
$\Theta(\boldsymbol{r})$ can be written as a sum over contributions of such vortex (or antivortex) sublattices $\Theta(\boldsymbol{r})=\sum_{i=1}^{N_v}s_i\theta(\boldsymbol{r}-\boldsymbol{r}_i)$, where $s_i=+$ ($s_i=-$) for vortices (antivortices) and $\boldsymbol{r}_i$ is the position of the $i$th sublattice with respect to the origin. Within each sublattice, the phase $\theta(\boldsymbol{r})$ can be expressed by summing the contributions of all vortices in the sublattice,
\begin{equation}
\theta(\boldsymbol{r})=\lim_{M\to\infty}\left[\sum_{m,n=-2M}^{2M}\Arg(\boldsymbol{r}-m\boldsymbol{\tau}_1-n\boldsymbol{\tau}_2) \mod 2\pi\right],
\label{eq:method}
\end{equation}
where $\boldsymbol{\tau}_i=q_i a_i\hat{\boldsymbol{\tau}}_i$ are the vectors spanning the MUC, composed of $q_1\times q_2$ atomic sites. Using complex variables $z=x+iy$, we have
\begin{equation} 
	\theta(z) =
	\Im\left\lbrace\Log\left[ i\vartheta_1\left(\frac{z}{\tau_2},-\frac{\tau_1}{\tau_2}\right)\right] 
		- \frac{2iz^2}{\tau_1\tau_2} \arctan\left(\frac{i\tau_1}{\tau_2}\right)\right\rbrace,
\end{equation}
where $\tau_i$ is the complex representation of the vector $\boldsymbol{\tau}_i$. 

It is important to note that, although the resulting function $\theta(\boldsymbol{r})$ admits the correct windings at the positions of the vortices, it is generally nonperiodic on the MUC. Therefore, using this summation for taking PBCs for a single MUC (a torus) is unsafe. 

\section{Lattice periodic gauge} 
We proceed by taking a gauge transformation that renders the order parameter and the hopping amplitudes periodic in the MUC $\mathbf{A}\rightarrow \mathbf{A} +\frac{1}{2}\nabla_{\boldsymbol{r}}\chi, 
\Delta\rightarrow \Delta e^{i\chi}$, $\psi_{\boldsymbol{r}s}\rightarrow e^{i\chi/2}\psi_{\boldsymbol{r}s}$. We note that the supercurrent $\mathbf{J}\propto\frac{1}{2}\nabla_{\boldsymbol{r}}\Theta-\mathbf{A}$ is periodic in the two magnetic lattice vectors $\boldsymbol{\tau}_i$ and thus $\int_{\boldsymbol{r}}^{\boldsymbol{r}+\boldsymbol{\tau}_i}\mathbf{J}\cdot\mathbf{d}\boldsymbol{\ell}$ is similarly doubly periodic. Therefore, we can always choose $\chi(\boldsymbol{r})$ so that the fields $\Theta'(\boldsymbol{r})=\Theta(\boldsymbol{r})+\chi(\boldsymbol{r})$ and $\int_{\boldsymbol{r}}^{\boldsymbol{r}+\boldsymbol{\tau}_i}\left(\mathbf{A}+\frac{1}{2}\nabla_{\boldsymbol{r}}\chi\right)\cdot\mathbf{d}\boldsymbol{\ell}$ are periodic (mod $2\pi$) on the lattice sites $\boldsymbol{r}_{m,n}=(m/q_1) \boldsymbol{\tau}_1 + (n/q_2) \boldsymbol{\tau}_2$. We now show that there exists a gauge that fulfills the conditions above for a MUC composed of $q\times  (q+1)$ atomic sites for which $q_2-q_1=1$. For a general vortex lattice, using the same notation as for $\Theta(\boldsymbol{r})$ above, we write $\chi(\boldsymbol{r})=\sum_{i=1}^{N_v}s_i\phi(\boldsymbol{r},\boldsymbol{r}_i) $ where $\phi(\boldsymbol{r},\boldsymbol{r}_i)$ is written in terms of complex variables as
\begin{equation}\label{eq:argument_lattice}
\begin{aligned}&\phi(z,z_i)=2\Re\left[ \frac{(z-z_i)^2}{\tau_1\tau_2} \arctan\left(\frac{i\tau_1}{\tau_2}\right) \right] 
\plus q\pi\Re\left(\frac{z^2}{\tau_1 \tau_2 } \right)\\
&\minus (q+1)\pi\frac{\Im^2\left( {z/\tau_2}\right)\Re\left({\tau_1/\tau_2}\right)}{\Im^2\left({\tau_1/\tau_2}\right)} 
\minus q\pi\frac{\Im^2\left( {z/\tau_1}\right)\Re\left({\tau_2/\tau_1}\right)}{\Im^2\left({\tau_2/\tau_1}\right)} \\ &
\plus \pi\frac{\Im\left(z/\tau_1\right)}{\Im\left(\tau_2/\tau_1\right)}\plus \left[2\pi\Re\left(\frac{z_i}{\tau_2} \right) \minus \pi\right]\frac{\Im\left(z/\tau_2\right)}{\Im\left(\tau_1/\tau_2\right)}.
\end{aligned}
\end{equation}		
The resulting phase function $\Theta'$ is now doubly periodic as required. Furthermore, integrating the supercurrent $\mathbf{J}(\boldsymbol{r})$ around the MUC reveals that 
\begin{equation}
0=\oint_{\textrm{MUC}}\mathbf{J}\cdot\mathbf{d}\boldsymbol{\ell}\propto N_w\Phi_0-\oint_{\textrm{MUC}}\mathbf{A}\cdot\mathbf{d}\boldsymbol{\ell},
\end{equation}
where $\Phi_0=h/(2e)=\pi$ is the superconducting magnetic flux quantum and $N_w=\sum^{N_v}_{i=1}s_i$ is the total winding for the vortices in the MUC. Due to the Dirac quantization condition [19], requiring that $\Phi=n(h/e)$ with $n\in\mathbb{Z}$ when taking PBCs on $\mathbf{A}$, $N_w$ must be an even number.

\section{The almost anti-symmetric gauge} Our next step is to find a complementary vector field. Due to the periodicity of the supercurrent, the vector potential is required to fulfill the condition,
\begin{equation}\label{eq:conditions}
\begin{aligned}
&\mathbf{A}(\boldsymbol{r}+\boldsymbol{\tau}_i)=\mathbf{A}(\boldsymbol{r}) + \frac{1}{2}\boldsymbol{\nabla}\left[ \Theta'(\boldsymbol{r}+\boldsymbol{\tau}_i)-\Theta'(\boldsymbol{r})\right].
\end{aligned}
\end{equation}
We now introduce the AAG that is designed to generate a homogeneous magnetic field and obey Eq.~(\ref{eq:conditions}) and is given by
\begin{equation} \label{eq:almost_lattice}
\mathbf{A}=\frac{2\Phi_0 p}{a_1 a_2 \sin ^2(\alpha_1 -\alpha_2 )} \left[\frac{(\boldsymbol{r} \times \hat{\boldsymbol{\tau}}_1)\times \hat{\boldsymbol{\tau}}_2}{ q+1}+\frac{  (\boldsymbol{r} \times\hat{\boldsymbol{\tau}}_2)\times \hat{\boldsymbol{\tau}}_1}{q}\right],
\end{equation}
where $\alpha_i=\Arg{\tau_i}$ and $p\in\mathbb{Z}\mod q(q+1)$. 

The AAG is also useful in other contexts. For example, if one is interested in solving the Hofstadter problem \cite{hofstadter1976energy} with high-flux resolution, it is obtained by considering a rectangular lattice of size $q \times  (q+1)$ and choosing an AAG $\mathbf{A}(\boldsymbol{r})=\nobreak 2\Phi_0 p\left(\frac{y}{q+1},\frac{x}{q}\right)$ with $p=1,2,\ldots,q(q+1)$. The flux per unit cell is then $\frac{2\Phi_0 p}{q(q+1)}$, and thus the flux through the entire 2D system is $2\Phi_0 p$. In the standard procedure using the Landau gauge, the flux through the entire 2D area can only take values from a narrow and sparse range $2\Phi_0 pq$ with $p=1,2,\ldots,q+1$.

\section{Electronic band structure of a vortex lattice}
\begin{figure}[!htb] 
	\centering
	\includegraphics[trim={0 0 0 0},clip,width=0.485\textwidth]{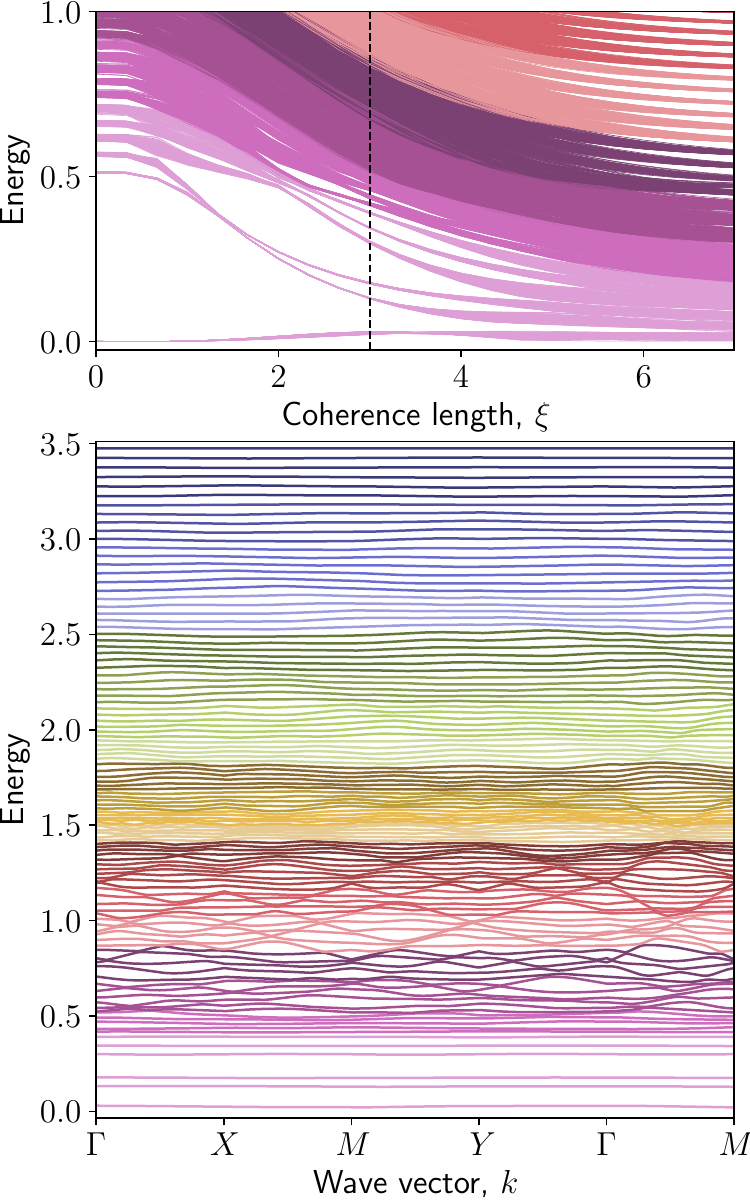}	
	\caption{(Top) Quasiparticle bands as function of coherence length $\xi$ for a pinned vortex lattice in a $p$-wave superconductor. The magnetic unit cell contains $10\times11$ sites and two vortices that are pinned on its diagonal, partitioning it in a ratio of 1:2:1. We take $t=|\Delta|=\mu=1$. (Bottom) The quasiparticle band structure for $\xi=2.5$. We observe Landau levels at high energies and Caroli-de Gennes-Matricon states below the gap, including the band generated from zero-mode tunneling \cite{grosfeld2006electronic}. \label{fig:bands}}
\end{figure} 
We now elucidate the quasiparticle energy dispersion for the pertinent BdG Hamiltonian, which is depicted in Fig.~\ref{fig:bands}. Consider a vortex lattice made of $N_1\times N_2$ MUCs with $q_1\times q_2$ atomic sites in each cell, so in total, the system consists of  $L_1\times L_2$ sites  ($L_i=N_i q_i$). The Hamiltonian of the vortex lattice in the BdG representation is written as $\hat{H} = \Psi^\dagger H_\mathrm{BdG} \Psi$, where $H_\mathrm{BdG}$ is the Hamiltonian density. For $s$-wave superconductors, $\Psi\equiv(\psi_\downarrow,\psi_\uparrow^\dagger)^T$  where $\psi_s$ with $s \in \{\uparrow, \downarrow\}$ is an $L_1 L_2$ component spinor of spin $s$ fermion annihilation operators. For $p$-wave superconductors, the index $s$ indicates particle and hole subspaces.

Next, we introduce the discrete translation operators along the two lattice directions, $i=1,2$,
\begin{equation}
T_i:\quad \psi_{\boldsymbol{r},s} \longrightarrow \psi_{(\boldsymbol{r}+\boldsymbol{\tau}_i)\,\mathrm{mod}\, N_i \boldsymbol{\tau}_i,s},
\end{equation}
which satisfy $[T_1,T_2]=0$ and $[H_\mathrm{BdG},T_i]=0$. Clearly, the eigenvalues of $T_i$ are $e^{i 2\pi n_i/N_i}$ with $n_i=1,2,\ldots,N_i$. The Bloch theorem is employed by introducing $q_1\times q_2$ sublattice wave functions,
\begin{eqnarray}
	\varphi_{\boldsymbol{k},s}(\boldsymbol{r})=\frac{1}{\sqrt{N_1 N_2}}\sum_{\boldsymbol{R}} e^{i \boldsymbol{k}\cdot\boldsymbol{R}}|\boldsymbol{R}+\boldsymbol{r},s\rangle,
\end{eqnarray}
where $\boldsymbol{R}\equiv\boldsymbol{R}_{m_1,m_2}=m_1\boldsymbol{\tau}_1+m_2\boldsymbol{\tau}_2$ denotes the positions of the MUCs and $\boldsymbol{k}\equiv\boldsymbol{k}_{n_1,n_2}=\frac{2\pi n_1}{N_1|\boldsymbol{\tau}_1|}\hat{\boldsymbol{\tau}}_1+\frac{2\pi n_2}{N_2|\boldsymbol{\tau}_2|}\hat{\boldsymbol{\tau}}_2$. 
The Hamiltonian within a given sublattice is defined as
\begin{eqnarray}
H_{\boldsymbol{k}}(\boldsymbol{r},s;\boldsymbol{r}',s')=\langle\varphi_{\boldsymbol{k},s}(\boldsymbol{r})|H_{\mathrm{BdG}}|\varphi_{\boldsymbol{k},s'}(\boldsymbol{r'})\rangle.
\end{eqnarray}
In this notation, the particle-hole symmetry of each block takes the form $\Sigma_1 H_{-\boldsymbol{k}}^* \Sigma_1 = -H_{\boldsymbol{k}}$ with $\Sigma_1=\sigma_1\otimes I_{q_1 q_2}$. The block $H_{\boldsymbol{k}=\boldsymbol{0}}$ corresponds to a single MUC with PBCs. Technically, $H_{\boldsymbol{k}}$ is obtained from $H_{\boldsymbol{0}}$ just by varying the boundary conditions as follows:
\begin{equation}\begin{aligned}
H_{\boldsymbol{0}}(\boldsymbol{r},s;\boldsymbol{r}+\boldsymbol{\tau}_i,s') \rightarrow H_{\boldsymbol{0}}(\boldsymbol{r},s;\boldsymbol{r}+\boldsymbol{\tau}_i,s') e^{-i \boldsymbol{k}\cdot \boldsymbol{\tau}_i}, 
\end{aligned}\end{equation}
for any $\boldsymbol{r}$ on the boundary of the MUC.

\section{The anomalous charge response function \texorpdfstring{$c_{xy}$}{}} 

\begin{figure}[!tb]
	\centering
	\includegraphics[trim={0 0 0 0},clip,width=0.485\textwidth]{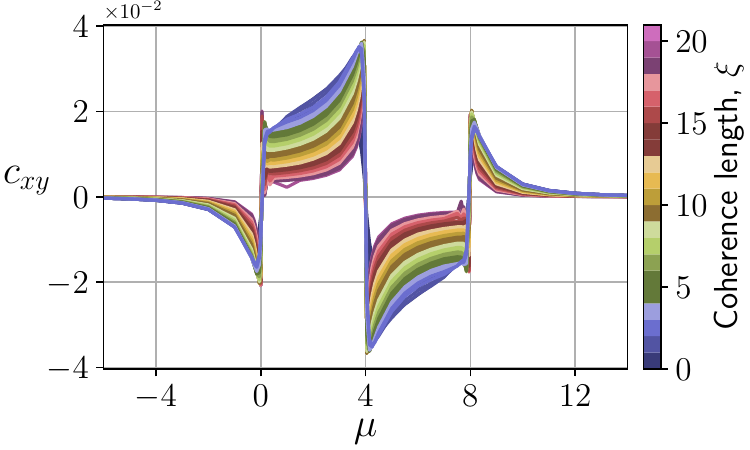}	
	\caption{Average anomalous charge response $c_{xy}$ vs chemical potential $\mu$ for different coherence lengths $\xi$. The $p$-wave superconductor has a magnetic unit cell of $40\times41$ sites $t=|\Delta|=1$. In addition, we pinned two vortices on the magnetic unit cell diagonal, partitioning it in a ratio of 1:2:1.\label{fig:xi}}
\end{figure} 

In previous studies of bulk $p$-wave superconductors, it was noted that $c_{xy}$ is not quantized \cite{goryo1999observation,stone2004edge,ariad2015effective}. We now calculate $c_{xy}$ in the presence of finite-size vortices and discover, remarkably, that $c_{xy}$ contains a universal quantized contribution.
 
The anomalous charge response is exposed in the effective action of a $p$-wave superconductor through  the appearance of a partial Chern-Simons (pCS) term \cite{roy2008collective,lutchyn2008gauge},
\begin{equation}
S_{\mathrm{pCS}}=\pm c_{xy} \int d\boldsymbol{r}dt~a_{t}\left(\nabla\times\boldsymbol{a}\right)_z,
\end{equation}
where $a_\mu=A_\mu-\partial_\mu\Theta/2$, $\mu\in\{t,x,y\}$ and the sign corresponds to the superconductor chirality $p_x\pm ip_y$. Thus, in analogy with the Streda formula \cite{streda1982theory}, the following relation holds  \cite{ariad2015effective}:
\begin{equation}\label{eq:streda}
c_{xy}(\boldsymbol{r}) = \pm \left.\frac{\partial \rho(\boldsymbol{r})}{\partial B_z}\right\vert_{B_z=0},
\end{equation}
where $\rho(\boldsymbol{r})=\delta S_{\mathrm{eff}}/\delta a_{t}(\boldsymbol{r})=\langle \text{gs} |\sum_s \psi^\dagger_{\mathbf{r},s}\psi_{\mathbf{r},s}|\text{gs}\rangle$, $|\text{gs}\rangle$ is the superconducting ground state and $B_z=\left(\nabla\times\boldsymbol{a}\right)_z$ is homogeneous at the lattice sites. This formula relates the density response to an infinitesimal external magnetic field. However, any variation of the magnetic field imposes a change in the superconducting phase in order to maintain periodicity of the supercurrents. Thus, as we now explain, the physical scenario here requires a modification of the Streda formula. The minimal variation of the magnetic field is a single flux quantum (over the entire system), leading to the nucleation of two vortices. Similarly, when an opposite magnetic field is applied, two antivortices are nucleated. Therefore, the derivative operation 
in the Streda formula for calculating density response implies a simultaneous  flip of magnetic field as well as vortex chiralities. This is equivalent to a chirality flip of the order parameter (from $p_x\pm ip_y$ to $p_x\mp ip_y$). The above procedure is also necessary as two opposite chirality states admit roughly the same spectrum so that the density response can be considered as a small perturbation. 

\begin{figure*}[!bt] 
	\centering
	\includegraphics[width=1\textwidth]{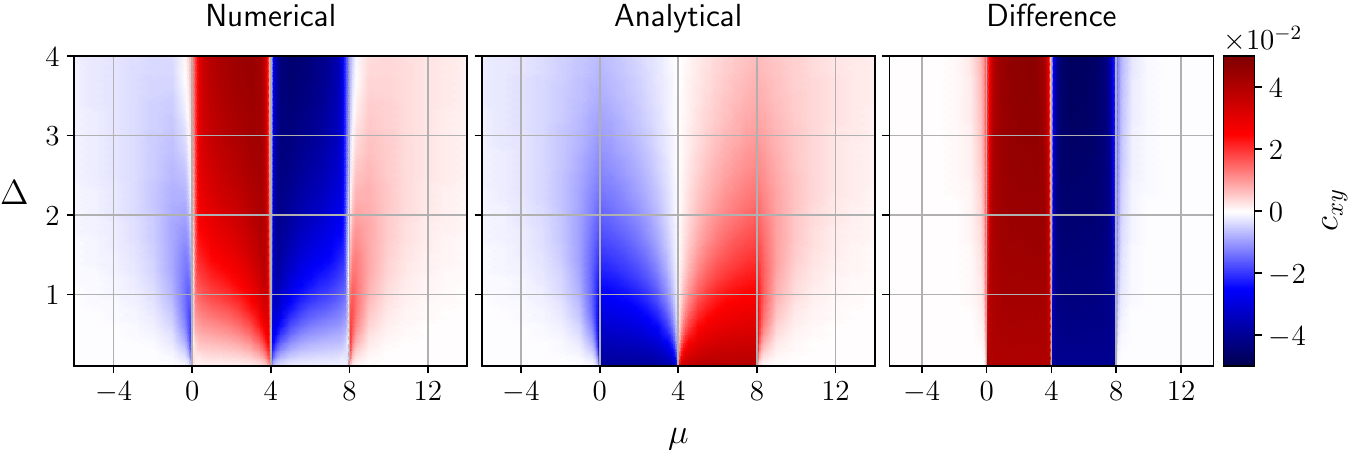}	
\caption{Average anomalous charge response $c_{xy}$ vs chemical potential $\mu$ and order parameter $|\Delta|$ for a $p$-wave superconductor with a magnetic unit cell of $40\times41$ sites $t=1$ and $\xi=2.5$. The modified Streda formula (Numerical) is compared with the prediction of the effective low-energy theory of the $p$-wave superconductor (Analytical). In addition, we pinned two vortices on the magnetic unit cell diagonal, partitioning it in a ratio of 1:2:1.\label{fig:delta} }
\end{figure*}

With this insight in mind, it is now possible to use Eq.~(\ref{eq:streda}) and numerically calculate the spatial average of $c_{xy}(\boldsymbol{r})$ as a function of $\mu$ as shown in Fig.~\ref{fig:xi}. The results are then compared with the analytical expression of $c_{xy}$ from the effective action governing the low-energy dynamics of the $p$-wave superconductor \cite{goryo1999observation,stone2004edge,ariad2015effective}.

It is found that the two predictions overlap in the trivial phases except that the numerics predict a slight dependence on $\xi$ but not on $|\Delta|$ as shown in Fig.~\ref{fig:delta}. Moreover, in all phases, $c_{xy}$ does not depend on the number of MUCs that form the vortex lattice. Hence, $c_{xy}$ can be calculated from a single MUC corresponding to $\boldsymbol{k}=\boldsymbol{0}$. Another property of $c_{xy}$ is that its average value within the MUC depends only slightly on its dimensions (as long as the vortices are well separated). Thus, one may expect to obtain $c_{xy}$ for $B_z=0$ by probing the density response of a small piece of the superconductor with PBCs for the application of minimal magnetic flux $\Phi=h/e$ and a compensating vortex pair (placed arbitrarily within the superconductor). This is indeed what we observe, and the result matches extremely well with the field-theoretical prediction in the trivial phase. Remarkably, in the topological phases ($0<\mu<8$) there is a sizable discrepancy between our predictions and those based on field theory. Since the charge accumulated at the vortex core (referred to as vortex charging) depends on the angular momentum of the Cooper pairs, it is determined by an interplay among the superconductor chirality, the vorticity and the quantum phase \cite{matsumoto2001vortex}. We now show that this discrepancy can indeed be traced to a universal vortex charging effect. 

\begin{figure}[!b] 
	\centering
	\includegraphics[trim={0 0 0 0},clip,width=0.475\textwidth]{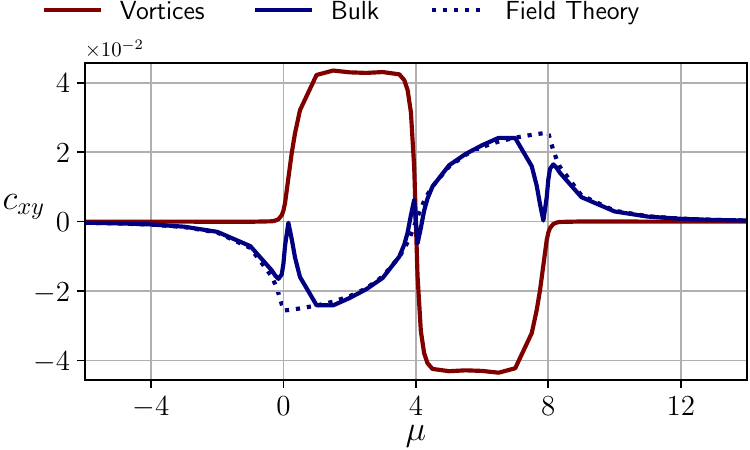}		
	\caption{Average anomalous charge
response $c_{xy}$ vs chemical potential $\mu$ for a planar $p$-wave superconductor. The magnetic unit cell average of $c_{xy}$ is crudely separated into contributions from the \textit{vortices} and contributions from the \textit{bulk}. For comparison, we also present the \textit{field-theory} prediction of $c_{xy}$. Here $t=|\Delta|=1$ and $\xi=2.5$. In addition, the magnetic unit cell contains $40\times41$ sites and two vortices that are pinned on its diagonal, partitioning it in a ratio of 1:2:1. \label{fig:separation}}
\end{figure}

To decipher the origin of $c_{xy}$, we perform two kinds of spatial and spectral cuts. First, we crudely separate the vortex cores at distances $r\leq\xi$ from the bulk and average $c_{xy}$ in each region independently to find their respective contributions; in the bulk, both theories yield similar results, whereas at the cores, the numerical results expose steps of $\pm\frac{1}{8\pi}$ as shown in Fig.~\ref{fig:separation}. Second, we separate the charge in the vortices into contributions of each Bogoliubov quasiparticle and take into account those within the energy gap, $\Delta Q_\text{core}=\iint_\text{core}\mathrm{d}\boldsymbol{r} \Delta\tilde{\rho}_{\boldsymbol{r}}$ with $\tilde{\rho}_{\boldsymbol{r}}=\frac{1}{2}\sum_{0<\epsilon< E_\text{gap}}\left( \vert v_{{\boldsymbol{r}},\epsilon}\vert^2 - \vert u_{{\boldsymbol{r}},\epsilon}\vert^2\right)$. We then find that the most significant contribution to $c_{xy}$ arises from the Caroli-de Gennes-Matricon states \cite{caroli1964bound}. This demonstrates that the universal contribution to $c_{xy}$ arises from the vortex core and, specifically, from vortex bound states. On the other hand, within the field theory  formalism,  the vortices are  treated as point singularities, which may explain the discrepancy. Altough it was observed in Ref.~\cite{matsumoto2001vortex} that vortices with opposite vorticities accumulate different charges, here we show that the relative accumulated charge for opposite vorticities is a universal quantity, which appears to be proportional to the Chern number of the superconductor. For consistency, we checked that $s$-wave and $d_{x^2-y^2}$-wave superconductors have vanishing anomalous charge responses.

\section{Summary} 
In this paper, the nature of the PKE and the order parameter in the $p\pm ip$ superconductor Sr$_2$RuO$_4$ is analyzed. A smooth gauge is introduced, that can be used in conjunction with Bloch's theorem to diagonalize BdG Hamiltonians for infinite superconductors in various periodic vortex states. The dispersion of quasiparticle energies for such vortex states with a finite vortex core size is calculated beyond previous numerical studies, and  the occurrence of midgap states is demonstrated as the size of the core is increased. 

Employing the same diagonalization algorithm, and modifying the Streda formula, the anomalous charge response $c_{xy}$ is calculated in the absence of vortices. The structure of $c_{xy}$ is then used to identify the quantum phases of the pertinent systems. Our results indicate that in $p$-wave superconductors subjected to PBCs, $c_{xy}$ is calculable by their response to an applied weak magnetic field and the nucleation of a vortex pair. On the other hand, the average value of $c_{xy}$ within the bulk is only weakly affected by the size of the vortices' core or their positions in the MUC. It is then reasonable to perceive that the discrepancy with results based on the field-theory approach to $p$-wave superconductors is attributed to vortex charging, which occurs only in vortices with  finite core radii. 

Finally, it is worth expressing our hope that the AAG introduced here and the ensuing diagonalization algorithm will serve as useful tools in the study of similar systems, such as the Hofstadter butterfly in the presence of disorder \cite{hofstadter1976energy}.

\begin{acknowledgments}
DA and EG acknowledge support from the Israel Science Foundation (Grant No. 1626/16) and the Binational Science Foundation (Grant No. 2014345). YA acknowledges support from the Israel Science Foundation (Grant No. 400/12). 
\end{acknowledgments}

%

\end{document}